\documentclass[aps,prb,twocolumn,a4paper,superscriptaddress]{revtex4-1}

\usepackage{graphicx,amsmath,amssymb,mathrsfs,array,longtable,delarray,color}

\begin{document}
\newcommand{\Vtg}{$V_{\text{tg}}$}
\newcommand{\Vbg}{$V_{\text{bg}}$}
\newcommand{\Etop}{$E_{\text{top}}$}
\newcommand{\Ebottom}{$E_{\text{bottom}}$}
\newcommand{\minor}{$\langle011\rangle$}
\newcommand{\major}{$\langle01\bar1\rangle$}
\newcommand{\modalpha}{$\lvert\alpha\rvert$}


\title{Orientation of hole quantum Hall nematic phases in an out-of-plane electric field}

\author{A.F.~Croxall}
\affiliation{Cavendish Laboratory, University of Cambridge, J.J.~Thomson Avenue, Cambridge CB3 0HE, United Kingdom}
\author{F.~Sfigakis}
\affiliation{Cavendish Laboratory, University of Cambridge, J.J.~Thomson Avenue, Cambridge CB3 0HE, United Kingdom}
\affiliation{Institute for Quantum Computing, University of Waterloo, 200 University Ave.~West, Waterloo, Ontario, Canada N2L 3G1}

\author{J.~Waldie}
\affiliation{Cavendish Laboratory, University of Cambridge, J.J.~Thomson Avenue, Cambridge CB3 0HE, United Kingdom}
\author{I.~Farrer}
\affiliation{Cavendish Laboratory, University of Cambridge, J.J.~Thomson Avenue, Cambridge CB3 0HE, United Kingdom}
\affiliation{Department of Electronic and Electrical Engineering, University of Sheffield, Sheffield, S1 3JD, United Kingdom}
\author{D.A.~Ritchie}
\affiliation{Cavendish Laboratory, University of Cambridge, J.J.~Thomson Avenue, Cambridge CB3 0HE, United Kingdom}

\date{\today}

\begin{abstract}
We present observations of an anisotropic resistance state at Landau level filling factor $\nu=5/2$ in a two-dimensional hole system (2DHS), which occurs for certain values of hole density $p$ and average out-of-plane electric field $E_\perp$. The 2DHS is induced by electric field effect in an undoped GaAs/AlGaAs quantum well, where front and back gates allow independent tuning of $p$ and $E_\perp$, and hence the symmetry of the confining potential. For $p\approx2\times10^{11}$~cm$^{-2}$ and $E_\perp \approx -2 \times10^{5}$~V/m, the magnetoresistance  along \major\ greatly exceeds that along \minor, suggesting the formation of a quantum Hall nematic or ``stripe'' phase. Reversing the sign of $E_\perp$ rotates the stripes by $90^{\circ}$. We suggest this behavior may arise from the mixing of the hole Landau levels and a combination of the Rashba and Dresselhaus spin-orbit coupling effects.
\end{abstract}

\maketitle


\section{Introduction}

High quality two-dimensional electron systems (2DESs) in perpendicular magnetic fields with half-filled Landau levels (LLs) display a rich variety of ground states arising from electron-electron interactions. In the lowest ($N = 0$) LL, at half filling of the spin up (filling factor $\nu=1/2$) or spin down states ($\nu = 3/2$), a compressible Fermi liquid of composite fermions is observed \cite{Halperin1993, Du1993, Willett1997}. Here, $\nu = nh/eB$, where $B$ is the magnetic field and $n$ is the carrier density. In the $N=1$ LL, at $\nu= 5/2$ and $7/2$, high quality 2DESs show an incompressible fractional quantized Hall (FQH) state \cite{Willett1987, Pan1999_52}. At half filling in the higher ($N\geq2$) LLs, the magnetoresistance can become highly anisotropic \cite{Lilly1999, Du1999}, suggesting a transition to a density-modulated ``stripe'' phase. Indeed, Hartree-Fock theory had predicted the existence of a unidirectional charge density wave state at half-filling of high LLs \cite{Koulakov1996, Fogler1996, Moessner1996}. More recent theoretical work \cite{Fradkin1999, Fradkin2000, Wexler2001, Cooper2002, Fradkin2010} suggests the stripe phase probably lacks long-range translational order and is more similar to a nematic phase \footnote{As well as the nematic, smectic and anisotropic crystal ground states are possible; see Q.~Qian, J.~Nakamura, S.~Fallahi, G.C.~Gardner, and M.J.~Manfra, Nature Commun.~\textbf{8}, 1536 (2017) and references therein.}. For GaAs-based 2DESs grown on the $(100)$ surface, in all but a few cases \cite{Zhu2002, Cooper2004, Liu2013, Pollanen2015} the stripes are found to align along the \minor\ direction, so that the magnetoresistance along \major\ is much higher than along \minor\ \cite{MacDonald2000}. Despite much investigation, it is still not clear what symmetry breaking mechanism causes the stripes to align in this way \cite{Cooper2001, Willett2001a, Pollanen2015, Shi2016a, Shi2016b, Shi2017a}. An externally applied symmetry breaking mechanism, such as an in-plane magnetic field \cite{Lilly1999a, Pan1999, Xia2010, Jungwirth1999, Stanescu2000} or periodic potential modulation \cite{Mueed2016} can re-orient the stripes along \major. In the $N=1$ LL, the FQH and stripe phases are thought to be very close in energy \cite{Rezayi2000}, and with an in-plane magnetic field the FQH states at $\nu=5/2$ and $7/2$ give way to an anisotropic state \cite{Lilly1999a, Pan1999,Xia2010}. Hydrostatic pressure can also bring about a transition from the $\nu=5/2$ and $7/2$ FQH states to a nematic phase \cite{Samkharadze2015, Schreiber2018}.

Two-dimensional hole systems (2DHSs) show somewhat similar behavior to 2DESs \cite{Shayegan2000, Manfra2006, Manfra2007, Takado2007, Koduvayur2011}. However, the FQH states at $\nu = 5/2$ and/or $7/2$ are often replaced by a stripe phase. This has been attributed to Landau level mixing caused by spin-orbit coupling \cite{Manfra2007} and/or hole-hole interactions, which become more prominent due to the large effective hole mass \cite{Shayegan2000}.

In this paper we present observations of an anisotropic state in a 2DHS in a GaAs quantum well (QW) at $\nu = 5/2$, which we believe to be a nematic/stripe phase. Our 2DHS is completely undoped, and symmetric front and back gates give us freedom to vary independently the 2DHS density and the asymmetry of the confining potential \cite{Croxall2013, Marcellina2018}. The anisotropic state occurs for a narrow range of densities at sufficiently low temperature. The orientation of the stripes is found to rotate by $90^\circ$ when the direction of the electric field perpendicular to the 2DHS plane is reversed. We suggest this behavior may arise from a combination of the Rashba and Dresselhaus spin-orbit coupling effects, as predicted in Ref.~\onlinecite{Sodemann2013}.

\section{Methods}
\begin{figure}
\centering
\includegraphics[width=0.48\textwidth]{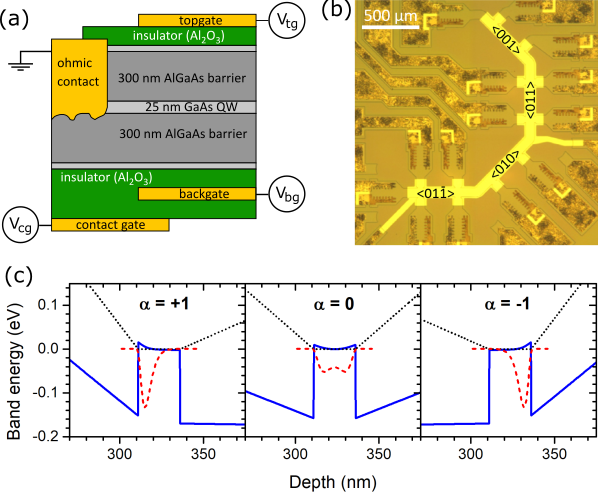}
\caption{(a) Schematic side view of the device. (b) Optical image of the top side of the device, before substrate removal and back-side processing. Labels indicate the crystallographic direction of hole transport in each section of the device. (c)-(e) Self-consistently calculated valence band edge (blue, solid), Fermi energy (black, dotted) and lowest subband wavefunction (red, dashed) for $p = 2\times10^{11}$~cm$^{-2}$ and (c) $\alpha = 1$, (d) $\alpha = 0$, and (e) $\alpha = -1$.}
\label{fig:Device}
\end{figure}

Our device, illustrated in Fig.~\ref{fig:Device}(a), contains a 2DHS induced by electric-field effect in a 25-nm-wide GaAs QW between two 300-nm-wide Al$_{0.33}$Ga$_{0.67}$As barriers. The wafer is grown by molecular-beam epitaxy on the $(100)$ GaAs surface. The device design and fabrication procedure are similar to those described in Ref.~\onlinecite{Croxall2013}. 
The semi-insulating substrate is completely removed, leaving a symmetric structure where the 2DHS density $p$ and the out-of-plane electric field $E$ can be varied independently using the voltages \Vtg\ and \Vbg\ applied to gates on the top and bottom surfaces of the device. We quantify the asymmetry of the confining potential by the dimensionless parameter $\alpha = (\text{\Etop}+\text{\Ebottom})/(\text{\Etop}-\text{\Ebottom})$, where $E_\text{top(bottom)}$ is the electric field in the barrier above/below the QW. The electric field difference $(\text{\Etop}-\text{\Ebottom})\propto p$, and we denote the average electric field by $E_\perp = (\text{\Etop}+\text{\Ebottom})/2$. Figures \ref{fig:Device}(c) to \ref{fig:Device}(e) illustrate the valence band profile and 2DHS wavefunction for $\alpha = 1$, $0$ and $-1$, calculated within an effective-mass approximation at zero magnetic field \footnote{Figures 1(c) to (e) are calculated using the nextnano++ software, see {https://www.nextnano.de/}}. For all the densities studied here, only the lowest subband of the QW is occupied \cite{Croxall2013}. The 2DHS mobility depends on $p$, $\alpha$, and direction, but is generally $>10^6$~cm$^2$/Vs.

The device geometry [Fig.~\ref{fig:Device}(b)] allows measurement of the resistivity $\rho$ along the \major, $\langle010\rangle$, \minor\ and $\langle001\rangle$ directions. Measurements are carried out in a dilution refrigerator with a base temperature of 50~mK, using standard low-frequency ac techniques. Unless otherwise specified, the excitation current is 1.3~nA or less, to limit electron heating.

\section{Results}
\begin{figure}
\centering
\includegraphics[width=0.48\textwidth]{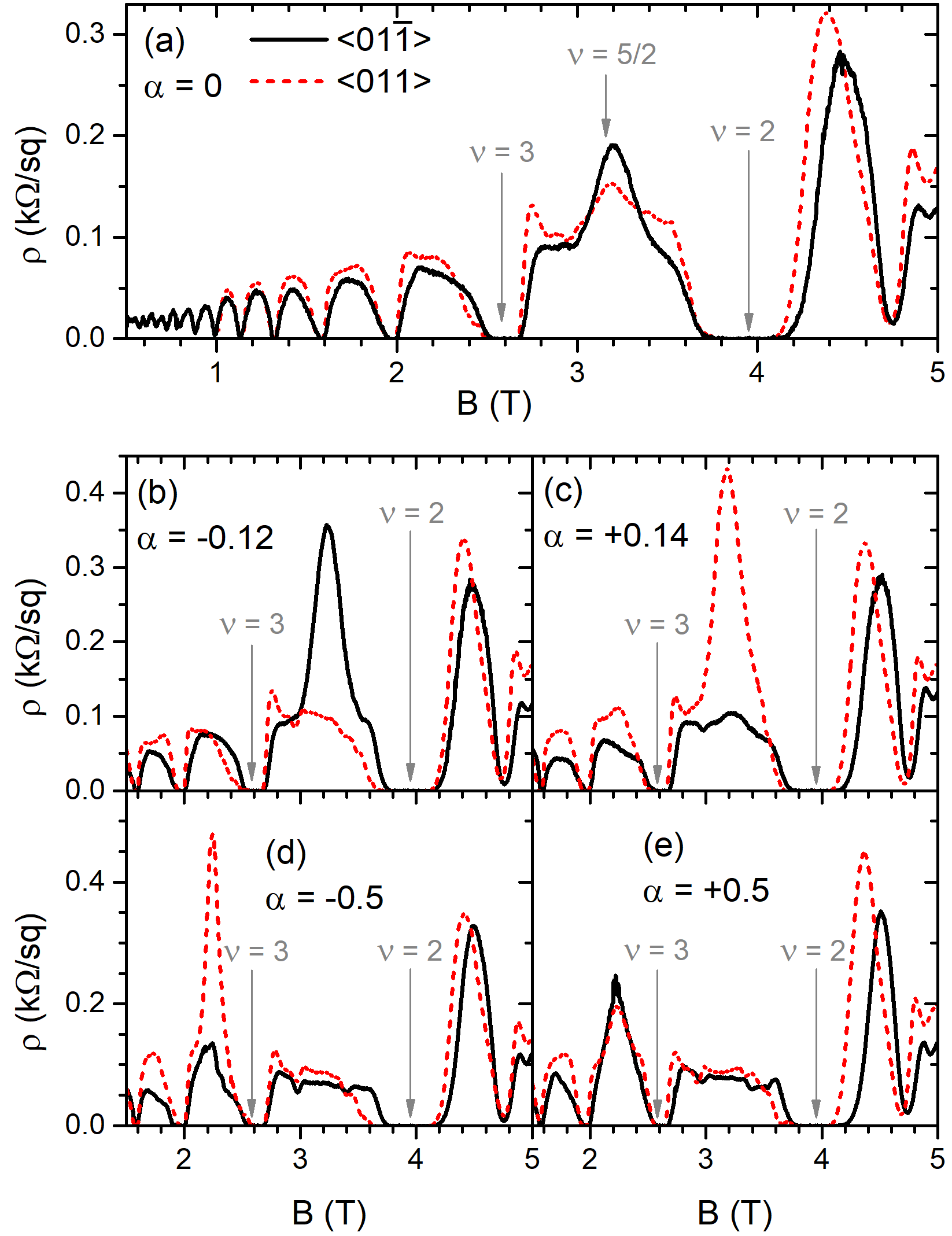}
\caption{2DHS resistivity in the \major\ (black, solid) and \minor\ (red, dashed) directions as a function of perpendicular magnetic field, for $p = 1.91\times10^{11}$~cm$^{-2}$, for (a) $\alpha = 0$, (b) $\alpha = -0.12$, (c) $\alpha = 0.14$, (d) $\alpha = -0.5$, and (e) $\alpha = 0.5$. All data are measured at $T = 50$~mK.}
\label{fig:Bsweeps}
\end{figure}

Our first experimental finding is that at $\nu = 5/2$, for certain densities and degrees of asymmetry, at low enough temperature, we see a large anisotropy in the 2DHS resistivity. Figure \ref{fig:Bsweeps}(a) shows the resistivity as a function of magnetic field $B$ in the \major\ and \minor\ directions, at $p = 1.91\times10^{11}$~cm$^{-2}$ when $\alpha = 0$. At $\nu = 5/2$ the resistivity in both directions is similar in magnitude. As $\alpha$ becomes increasingly positive(negative) the peak in the \minor(\major) direction increases while the peak in the \major(\minor) direction diminishes. This leads to a large resistivity anisotropy, which is maximized for $\lvert\alpha\rvert \approx 0.10$-$0.15$ [Figs.~\ref{fig:Bsweeps}(b) and \ref{fig:Bsweeps}(c)]. As $\lvert\alpha\rvert$ is further increased, the resistivity peak in the high resistance direction diminishes, so that isotropic behavior is restored for $\lvert\alpha\rvert \gtrsim 0.5$ and signs of FQHSs appear at $\nu = 7/3$ and $8/3$ [Figs.~\ref{fig:Bsweeps}(d) and \ref{fig:Bsweeps}(e)]. We note that in the anisotropic state, where the resistivity shows a maximum in one direction, the resistance in the orthogonal direction does not show a minimum, as is often observed in quantum Hall nematic states. This may be because our device resembles a Hall bar, rather than a van der Pauw geometry. The van der Pauw geometry has been shown to accentuate the degree of anisotropy by current channelling effects \cite{Simon1999, Willett2001}.

\begin{figure}
\centering
\includegraphics[width=0.48\textwidth]{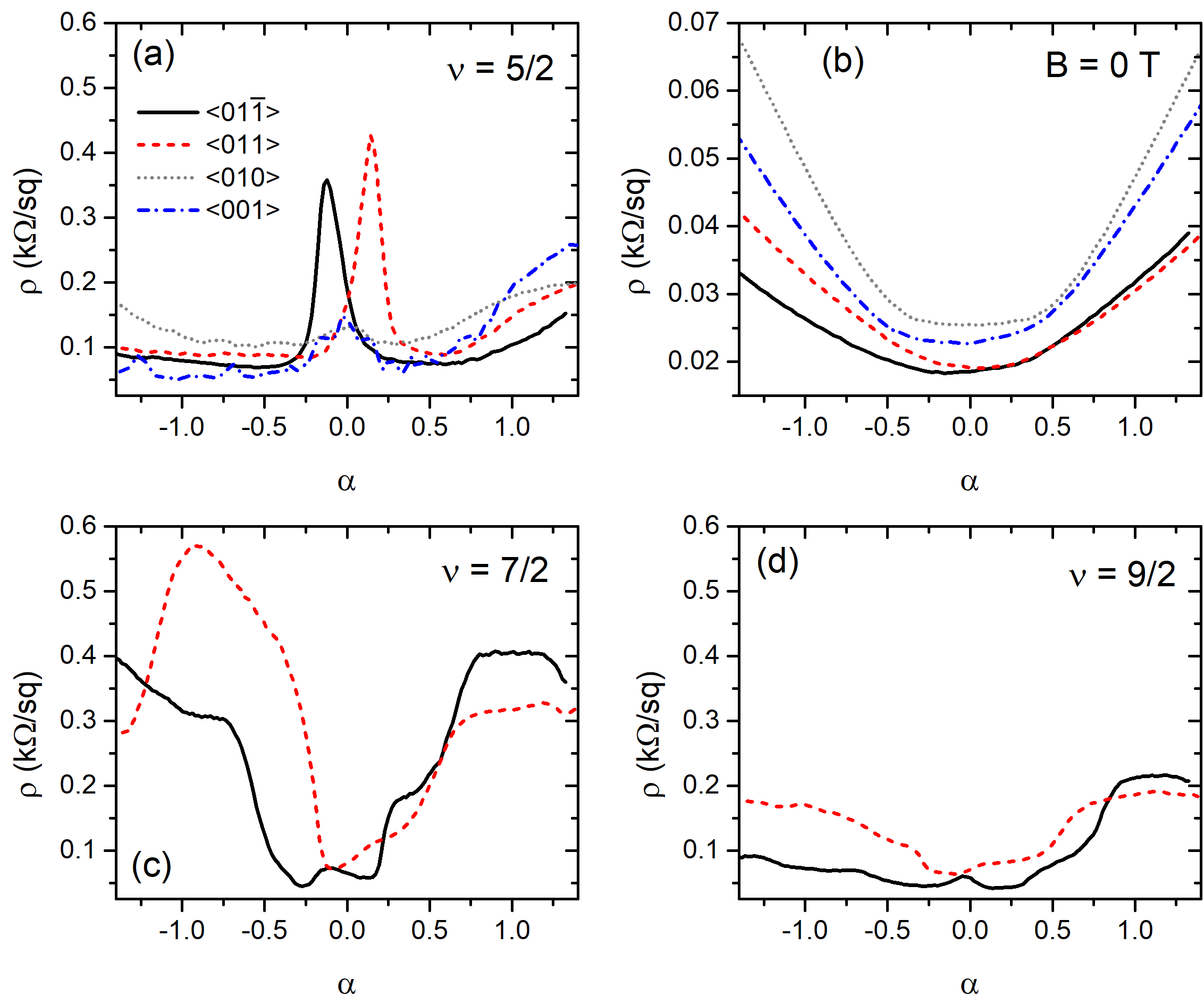}
\caption{2DHS resistivity as a function of $\alpha$ in the \major\ (black, solid), \minor\ (red, dashed), $\langle010\rangle$ (grey, dotted), and $\langle001\rangle$ (blue, dash-dot) directions, at (a) $\nu = 5/2$ ($B = 3.20$~T), (b) $B = 0$~T, (c) $\nu = 7/2$ ($B = 2.22$~T), and (d) $\nu = 9/2$ ($B = 1.73$~T). The temperature is 50~mK and the 2DHS density is $1.91\times10^{11}$~cm$^{-2}$. The data in (b) are taken with excitation current 5~nA.}
\label{fig:alpha_sweeps}
\end{figure}

Figure \ref{fig:alpha_sweeps}(a) summarizes the dependence of the resistivity in all four directions on $\alpha$ at $\nu=5/2$, for the same density as Fig.~\ref{fig:Bsweeps}. In contrast to the \major\ and \minor\ directions, we do not see any significant peak in $\rho(\alpha)$ along the $\langle001\rangle$ or $\langle010\rangle$ directions, in agreement with previous results \cite{Manfra2006}. For all four directions there is a general trend of increasing resistivity as \modalpha\ increases beyond approximately 0.5, both at $\nu = 5/2$ [Fig.~\ref{fig:alpha_sweeps}(a)] and at $B = 0$~T [Fig.~\ref{fig:alpha_sweeps}(b)]. This may be related to increased scattering from GaAs/AlGaAs interface roughness disorder as the 2DHS wavefunction is pushed towards the edges of the QW at large \modalpha\ [see Figs.~\ref{fig:Device}(c) and \ref{fig:Device}(e)] \cite{Croxall2013}. However, the sharp peaks in $\rho(\alpha)$ along the \major\ and \minor\ directions, for small negative and positive $\alpha$ respectively, have no counterpart at $B = 0$~T.

We have looked for similar effects at $\nu = 7/2$ [Fig.~\ref{fig:alpha_sweeps}(c)] and $\nu = 9/2$ [Fig.~\ref{fig:alpha_sweeps}(d)]. At $\nu = 9/2$, for all values of $\alpha$ we have studied, there is no evidence of anisotropy significantly greater than that at $B = 0$~T. The behavior at $\nu = 7/2$ is complex. Both $\rho_{\text{\major}}$ and $\rho_{\text{\minor}}$ show a broad minimum at small $\lvert\alpha\rvert$, flanked by peaks at $\lvert\alpha\rvert \sim 1$. It may be significant that the peaks in $\rho(\alpha)$ for $\nu = 5/2$ occur at roughly the center of the broad minima in $\rho(\alpha)$ for $\nu = 7/2$. However, at $\nu = 7/2$, for $-1.2 \lesssim \alpha \lesssim -0.2$, $\rho_{\text{\minor}}$ is significantly greater than $\rho_{\text{\major}}$ [see also Fig.~\ref{fig:Bsweeps}(d)], while, for positive $\alpha$, $\rho_{\text{\minor}}$ and $\rho_{\text{\major}}$ are very similar in magnitude. While these observations are suggestive of stripe ordering at $\nu = 7/2$ at certain $\alpha$, further investigations are required to determine the nature of this state.

\begin{figure}[b]
\centering
\includegraphics[width=0.48\textwidth]{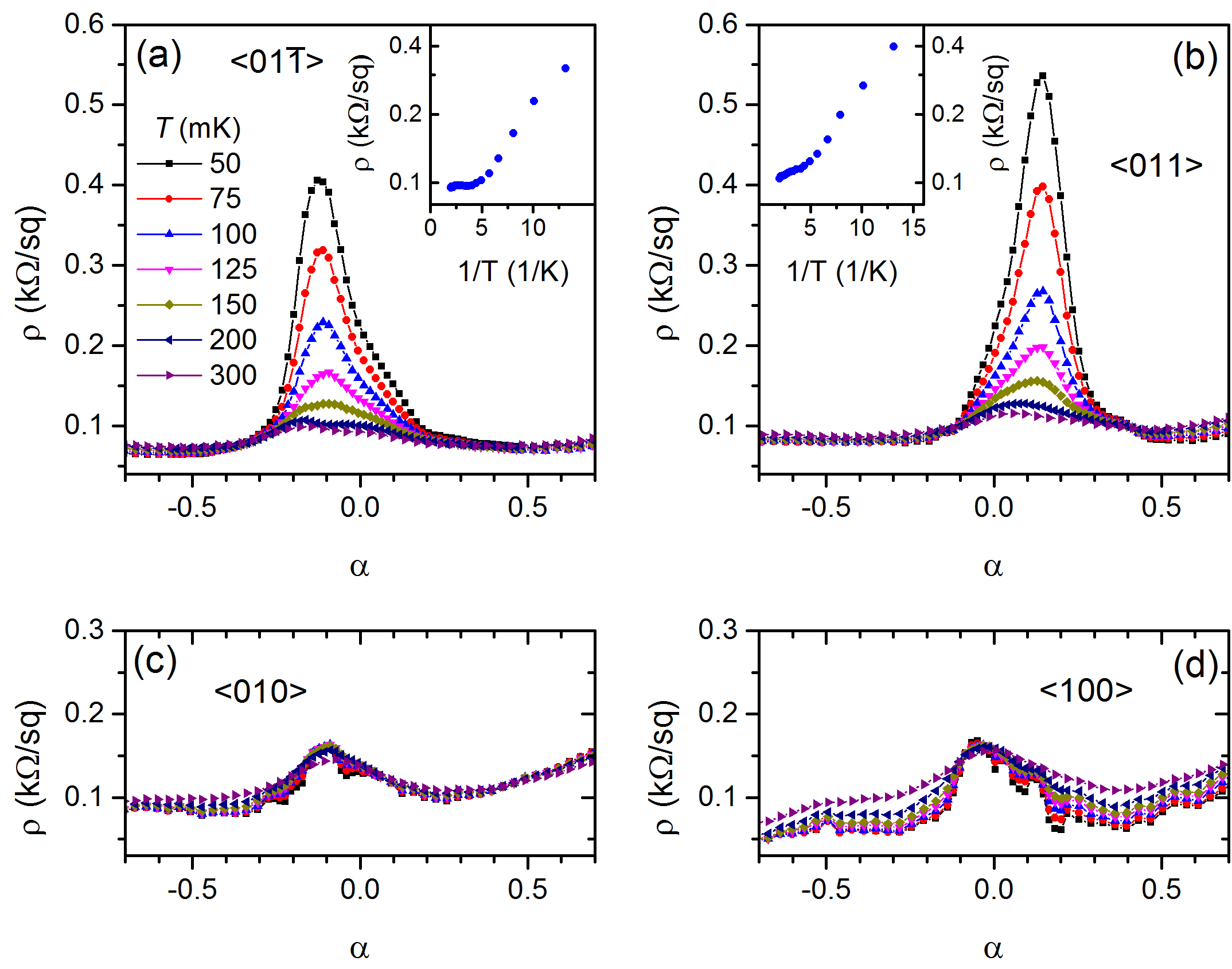}
\caption{2DHS resistivity at $\nu = 5/2$ as a function of $\alpha$ at different temperatures, in the (a) \major, (b) \minor, (c) $\langle010\rangle$, and (d) $\langle001\rangle$ directions. The 2DHS density is $2.04\times10^{11}$~cm$^{-2}$ and $B = 3.38$~T. The insets in (a) and (b) show the peak resistivity (on a logarithmic scale) versus reciprocal temperature for the \major\ and \minor\ directions respectively.}
\label{fig:Tdependence}
\end{figure}

We now consider the temperature dependence of the resistivity in the anisotropic state at $\nu = 5/2$. In the $\langle001\rangle$ and $\langle010\rangle$ directions, $\rho$ is only weakly temperature dependent up to 300~mK [Figs.~\ref{fig:Tdependence}(c) and ~\ref{fig:Tdependence}(d)]. The same is true in the \major\ and \minor\ directions for $\text{\modalpha} \gtrsim 0.3$, where the resistivity is isotropic [Figs.~\ref{fig:Tdependence}(a) and ~\ref{fig:Tdependence}(b)]. However, the temperature dependence of the peak in $\rho_{\text{\major}}$($\rho_{\text{\minor}}$) for small negative(positive) $\alpha$ is insulating ($d\rho/dT < 0$). In common with other studies of quantum Hall nematic states \cite{Lilly1999, Du1999, Wexler2001, Cooper2002}, the anisotropic state is destroyed by increasing temperature; in our case the resistance is isotropic for $T\gtrsim 200$~mK. Below this temperature, we find the peak resistivity follows $\rho \propto \rho_0 \exp{(T_0/T)}$ [see insets to Figs.~\ref{fig:Tdependence}(a) and ~\ref{fig:Tdependence}(b)], with characteristic temperature scale $T_0 \sim 160$~mK for both \major\ and \minor\ at $p = 2.04\times10^{11}$~cm$^{-2}$. We note that at $\nu = 9/2$ the resistivity in all directions is only weakly temperature dependent, while at $\nu = 7/2$ the broad peaks in $\rho_{\text{\minor}}$ and $\rho_{\text{\major}}$ around $\lvert\alpha\rvert \sim 1$ have an insulating temperature dependence similar to that of the resistivity peaks at $\nu = 5/2$, suggestive of stripe behavior.

\begin{figure}
\centering
\includegraphics[width=0.48\textwidth]{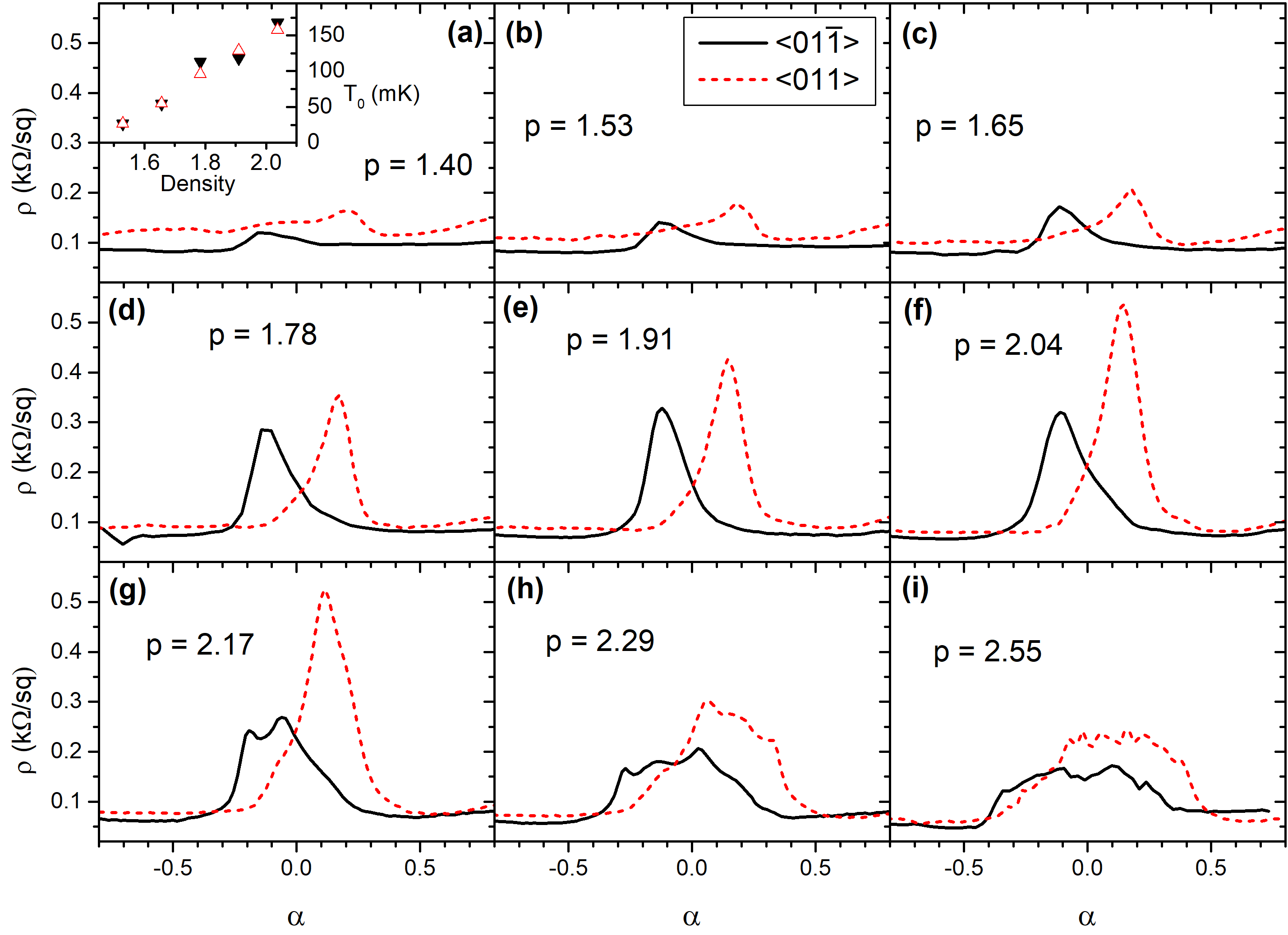}
\caption{Resistivity at $\nu = 5/2$  and $T=50$~mK as a function of $\alpha$ in the \major\ (black, solid) and \minor\ (red, dashed) directions for 2DHS densities (a) 1.40, (b) 1.53, (c) 1.65, (d) 1.78, (e) 1.91, (f) 2.04, (g) 2.17, (h) 2.29 and (i) 2.55 (units $10^{11}$~cm$^{-2}$). The inset in (a) shows the characteristic temperature $T_0$ of the increase of peak resistance with temperature for \major\ (filled downward triangles) and \minor\ (open upward triangles) as a function of density.}
\label{fig:density}
\end{figure}

In Fig.~\ref{fig:density} we explore the density dependence of the resistivity anisotropy at $\nu = 5/2$, comparing $\rho_{\text{\major}}(\alpha)$ and $\rho_{\text{\minor}}(\alpha)$ at a range of densities. For all densities there is a peak in $\rho_{\text{\major}}$($\rho_{\text{\minor}}$) centered at small negative(positive) $\alpha$. These peaks are strongest for $p\approx 2\times10^{11}$~cm$^{-2}$. At lower density the peaks diminish, while at higher density the peaks become broader so that the degree of anisotropy is reduced. In all cases the temperature dependence of the peak resistance is insulating. For $1.5\times10^{11}$~cm$^{-2} \lesssim p \lesssim 2.2\times10^{11}$~cm$^{-2}$ we again find $\rho \propto \rho_0 \exp{(T_0/T)}$, with $T_0$ increasing linearly with density [see inset to Fig.~\ref{fig:density}(a)]. This may be related to the increase of the magnetic field, and hence the cyclotron energy, at $\nu = 5/2$ as the density is increased \cite{Stanescu2000}. At higher density the temperature dependence of the resistivity peaks is more similar to a power law, $\rho \propto \rho_0 (T_0/T)^\beta$, with $0 < \beta < 0.5$, so that we cannot identify a characteristic temperature scale.

It is not obvious whether the broad peaks in $\rho(\alpha)$ at high density arise from a nematic state, since the resistivity is approximately isotropic. However, given the continuous evolution of these peaks from the anisotropic state at $p\approx 2\times10^{11}$~cm$^{-2}$, and their insulating temperature dependence, we suggest the nematic state may persist for $p> 2\times10^{11}$~cm$^{-2}$. For 2DES quantum Hall nematics at $\nu = 9/2$, it is thought that the states with density modulations (``stripes'') along \major\ and \minor\ are close in energy and that in certain circumstances domains of each orientation can co-exist \cite{Fil2000, Cooper2004, Zhu2009}. In our 2DHS at $\nu = 5/2$ there may be domains of both orientations, with the dominant orientation controlled by both density and $\alpha$. While the nematic orientation seems to weaken at low density, and be stronger in one ($\alpha$-dependent) direction than another for $p \sim 2\times10^{11}$~cm$^{-2}$, at higher density there could be a significant fraction of domains oriented in each direction.

Finally, we note that we could not find any signs of nematic ordering/anisotropic resistivity for 2DHSs in narrower QWs (10~nm and 15~nm), although this could be because of the lower mobility in narrower wells due to increased interface roughness scattering.

\section{Discussion}

Our results raise two questions: (1) Why does the nematic phase form in the 2DHS at $\nu = 5/2$ only for small but non-zero $\alpha \propto E_\perp$ and certain hole densities? (2) What is the symmetry breaking mechanism that orients the nematic and rotates it by $90^{\circ}$ when $E_\perp$ is reversed?

In 2DESs, nematic phases in a purely perpendicular magnetic field are usually only observed at half filling in the $N \geq 2$ LLs, in agreement with theoretical studies \cite{Fogler1997}. In the $N = 1$ LL at $\nu = 5/2$ and $7/2$ the nematic state can be stabilised by an in-plane magnetic field \cite{Lilly1999a, Pan1999,Jungwirth1999, Rezayi2000}. Observations of nematic states without an in-plane field at $\nu = 5/2$ and/or $7/2$ in 2DESs at very low density or under hydrostatic pressure \cite{Pan2014, Samkharadze2015, Schreiber2018} have been attributed to strong interaction-driven LL mixing when the ratio $\kappa = (e^2 / \epsilon\l_B) / \hbar\omega_c$ of the electron-electron interaction energy to the LL spacing is large. Here, $\l_B = (\hbar e/B)^{1/2}$ is the magnetic length, $\epsilon$ is the permittivity, and $\omega_c = eB/m^{*}$ is the cyclotron frequency. Stripe states have been observed at $\nu = 3/2$ in tilted magnetic field in systems where the Zeeman energy is large compared to the cyclotron energy, so that the chemical potential lies in the $N = 1$ LL \cite{Zhang2017, Hossain2018}. Conversely, in wide QWs the nematic can be absent at $\nu = 9/2$ because the chemical potential lies in, or close to, the $N = 0$ LL of the second QW subband \cite{Pan2000, Xia2010}.

Two-dimensional hole systems in GaAs/AlGaAs often show nematic phases at $\nu = 5/2$ and/or $7/2$ \cite{Shayegan2000, Manfra2006, Manfra2007, Takado2007, Koduvayur2011}. The larger effective mass in the 2DHS reduces the cyclotron energy, significantly enhancing the LL mixing parameter $\kappa$. Holes in GaAs are also subject to significant spin-orbit coupling \cite{Winkler}. Manfra \textit{et al.} observed an anisotropic state at $\nu = 7/2, 11/2$ and $13/2$ (but not at $\nu = 5/2$ or $9/2$) in a 2DHS in a symmetrically-doped 20-nm-wide QW with $p = 2.3\times10^{11}$~cm$^{-2}$, but not in samples with lower density or a strongly asymmetric confining potential \cite{Manfra2007}. This was attributed to strong spin-orbit coupling that mixes the valence band states and alters the orbital structure of the hole LLs at $B\neq0$ \cite{Yang1985, Ekenberg1985}, thus modifying the effective hole-hole interaction potential  \cite{MacDonald1989, Yang1990}. By self-consistent calculations of the LL structure, Manfra \textit{et al.} showed that their observations of a nematic phase correlated with the half-filled LL containing mostly $N \geq2$ orbitals, while isotropic states occurred for LLs containing significant amounts of the $N=0$ and $1$ orbitals. The LL mixing due to spin-orbit coupling depends strongly on the 2DHS density, the shape and width of the QW confining potential, and magnetic field \cite{Winkler}, so these factors will affect whether the anisotropic phase can form at various half-integer filling factors \cite{Manfra2007}.

We suggest the nematic phase at $\nu = 5/2$ in our 2DHS occurs because of Landau level mixing caused by both interactions and spin-orbit coupling. Both mixing mechanisms will be sensitive to the out-of-plane electric field, which may explain why the nematic is stable only at certain $\alpha$. The Rashba spin-orbit coupling term is proportional to $E_{\perp}$. As illustrated in Figs. \ref{fig:Device}(c) to \ref{fig:Device}(e), increasing $\lvert E_{\perp}\rvert$ effectively decreases the QW width. While decreasing QW width may enhance the hole-hole interaction, it will also increase the QW subband spacing and hence suppress mixing of the valence bands. This could be the reason why the nematic disappears for large $\lvert E_\perp \rvert$. However, these factors cannot account for what sets the orientation of the nematic ordering, especially the rotation of the stripes by $90^{\circ}$ when $E_\perp$ is reversed.

It is thought that the energy of the nematic state has local minima for orientation along \major\ or \minor\ \cite{Fil2000, Cooper2004, Zhu2009}, but which of these has the lowest energy is found to depend on factors such as density \cite{Zhu2002, Cooper2004}, 2DES depth \cite{Pollanen2015}, strain \cite{Koduvayur2011}, filling factor/spin \cite{Liu2013} and the orientation and magnitude of an in-plane magnetic field \cite{Pan1999, Lilly1999a, Shi2016a, Shi2016b, Shi2017a}. For unstrained samples on GaAs (100) surfaces, in a purely out-of-plane magnetic field, the \major\ and \minor\ directions are expected to be equivalent. We now discuss several proposed symmetry-breaking mechanisms and whether they are likely to be relevant in our system.

Koduvayur \textit{et al.} observed nematic states in a 2DHS at $\nu = 5/2$ and $7/2$ subjected to an in-plane shear strain $\varepsilon$, where the high resistance direction was rotated by $90^{\circ}$ upon reversing the sign of $\varepsilon$ (Ref.~\onlinecite{Koduvayur2011}).
These effects were explained by a strain-induced anisotropy of the exchange interaction, which preferentially orients the nematic density modulation along either \major\ or \minor\ depending on the sign of $\varepsilon$. Koduvayur \textit{et al.} argued that, since GaAs is a piezoelectric material, an out-of-plane electric field results in an in-plane strain, so the orientation of the nematic phases in unstrained devices could be related to asymmetries in the confining potential \cite{Koduvayur2011}. The experiments of Ref.~\onlinecite{Pollanen2015} showed no effect of the confining potential asymmetry on the nematic orientation in GaAs 2DESs. However, the predicted strain-induced anisotropy of the Harteee-Fock energy is two orders of magnitude larger for holes than electrons \cite{Koduvayur2011}. In our samples the anisotropy is maximized for average electric fields $\lvert E_\perp\rvert < 10^4$~V/cm, resulting in $\lvert\varepsilon\rvert < 3\times10^{-6}$ (using $\varepsilon = d_{14}E_{\perp}$, where $d_{14} = -2.7\times10^{-10}$~cm/V, following Ref.~\onlinecite{Koduvayur2011}). This is two orders of magnitude smaller than the strains applied by Koduvayur \textit{et al.}, so we do not think piezoelectric strain can account for our findings. Although we do not intentionally strain our samples, the fabrication technique of sample thinning before depositing the back gates could lead to residual strain. However, this strain would presumably be only weakly affected by the applied electric field, so it is difficult to see how it could lead to nematic phases in the \major\ or \minor\ sections of our Hall bar with different orientations at different electric fields.

Sufficiently strong periodic potential modulations can affect the stripe orientation \cite{Mueed2016, Yoshioka2001}. It is possible that the interface roughness on opposite sides of our quantum well has corrugations that pin the stripes in one direction when the electric field pushes the wavefunction to one side of the well and in the orthogonal direction for the opposite electric field. However, we consider it unlikely that surface corrugations during MBE growth could rotate by $90^\circ$ during just 25~nm of GaAs growth, or be so different in areas of the sample less than 1~mm apart, and  anisotropic interface roughness in MBE-grown 2DESs has been shown experimentally to have no consistent effect on the nematic orientation \cite{Willett2001a, Cooper2001}.

Sodemann and MacDonald have argued that the combination of the Rashba and Dresselhaus spin-orbit interactions breaks rotational symmetry and could be responsible for orienting the nematic states of GaAs/AlGaAs 2D electron systems along the \minor\ directions in most cases \cite{Sodemann2013}. Individually the Rashba and Dresselhaus effects would maintain rotational invariance, but this is broken by the two in combination. The predicted anisotropic contribution to the energy is
\begin{equation}
E^{\text{ani}}(\theta,a,\nu) = N_{\phi}\sin (2\theta) \varepsilon(a,\nu) \frac{\gamma_R \gamma_D}{(\omega_c l_B)^2} \frac{e^2}{\epsilon l_B},
\label{eqn:E_ani}
\end{equation}
where $\theta$ is the angle between the stripes and the $[001]$ direction, $a$ is the stripe period, $N_\phi$ is the orbital LL degeneracy, $\gamma_{R(D)}$ is the coefficient of the Rashba(Dresselhaus) term in the spin-orbit Hamiltonian  \footnote{In a 2DES, the spin-orbit Hamiltonian can be written $H_{\text{SO}} = \gamma_{R}(\sigma_x \pi_y - \sigma_y \pi_x) + \gamma_{D}(\sigma_y \pi_y - \sigma_x \pi_x)$, where $\pi$ is mechanical momentum and $\sigma$ is spin.} and $\varepsilon(a,\nu)$ is a dimensionless number. The Rashba coefficient  $\gamma_R$ is strongly dependent on $E_{\perp}$. Since $\gamma_R$ is an odd function of $E_{\perp}$, while $\gamma_D$ is even, the stripes are expected to rotate by $90^\circ$ when the sign of the electric field is reversed. This theory was developed for 2D electron systems. Further theoretical work would be needed to determine whether a similar mechanism could operate in a 2DHS, where the functional forms of the Rashba and Dresselhaus terms are somewhat different and there is much stronger LL mixing. However, the symmetry of the Rashba and Dresselhaus terms with respect to $E_{\perp}$ should be the same for both electrons and holes \cite{Winkler2000}, so we suggest it could explain our results. For example, our observation of the nematic phase for only a narrow range of $\lvert E_{\perp}\rvert$ could be because anisotropy is maximised when the Rashba and Dresselhaus terms are of similar strength.

We note that Pollanen \textit{et al.} made a detailed study of the effects of heterostructure asymmetry on the orientation of the 2DES nematic phase at $\nu = 9/2$ (Ref.~\onlinecite{Pollanen2015}). By varying the doping profile in quantum well structures, they found that the hard axis of the $\nu=9/2$ nematic state was always along the \major\ direction, independent of the sign of the average perpendicular electric field at the location of the 2DES \cite{Pollanen2015}. This implied that the spin-orbit model of Ref.~\onlinecite{Sodemann2013} is not a dominant symmetry breaker in 2DESs. However, in GaAs the spin-orbit coupling is much stronger for holes than for electrons, so we do not think the findings of Pollanen \textit{et al.} rule out a spin-orbit origin for our results.

Takhtamirov and Volkov \cite{Takhtamirov2000} and Rosenow and Scheidl \cite{Rosenow2001} have shown that the asymmetric confinement potential in a GaAs/AlGaAs heterostructure leads to anisotropy in the electron effective mass even without spin-orbit coupling, which may explain the preferential stripe orientation along \minor\ in GaAs 2DESs. This mechanism is also expected to be stronger for holes than for electrons \cite{Takhtamirov2000}. However, we might expect the effective mass anisotropy, and presumably the stripe stability, to increase with increasing $\lvert E_\perp\rvert$, while we find the anisotropy is destroyed beyond a critical $\lvert E_\perp\rvert$.

We must point out that our calculation of $E_\perp$ contains significant uncertainty because the Fermi-level pinning on opposite sides and in different regions of the sample may differ slightly, and because the QW subband energies vary as a function of $p$ and $E_\perp$. There is a small possibility that the peaks in resistance at $\nu = 5/2$ could be occurring at $E_\perp=0$. However, it is very unlikely that our error in calculating $E_\perp$ could be so different in different parts of the sample, or that the high resistance direction could be different in different parts of the sample if both parts of the sample really had $E_\perp=0$. Different sample geometries (e.g. van der Pauw) could be useful in ruling out such effects because the resistance for the \major\ and \minor\ directions could be measured in the same area of the sample.

\section{Conclusion}

In conclusion, we have reported observations of a quantum Hall nematic state in a 2DHS in a GaAs QW at $\nu = 5/2$ for a small, non-zero average perpendicular electric field $E_\perp$, of typical magnitude $\approx 2 \times10^{5}$~V/m. The nematic orientation rotates by $90^\circ$ when the direction of $E_\perp$ reversed. This behavior may be related to the mixing of the hole Landau levels under the combined action of the Rashba and Dresselhaus spin-orbit coupling effects.

\section*{Acknowledgements}
We thank I.~Sodemann, A.H.~MacDonald, A.R.~Hamilton and N.R.~Cooper for useful discussions. This work was supported by UK Engineering and Physical Sciences Research Council projects EP/H017720/1 and EP/J003417/1. A.F.C acknowledges funding from Trinity College at the University of Cambridge, UK. J.W acknowledges funding from the Herchel Smith Fund at the University of Cambridge. I.F. acknowledges funding from Toshiba Research Europe Limited.

The data presented in this article can be accessed at {https://doi.org/10.17863/CAM.38773}.

%

\end{document}